\documentclass{emulateapj}

\usepackage{subfigure}
\usepackage{url}
\usepackage{datetime}
\usepackage{longtable}
\usepackage{natbib}
\usepackage{amsmath}
\usepackage{ulem}
\usepackage{bm}
\usepackage{comment}

\newcommand{\noprint}[1]{}

\usepackage{color}

\newcommand{\itk}{{\it Kepler}}
\newcommand{\spitz}{{\it Spitzer}}

\newcommand{\mjup}{{M$_\textrm{Jup}$}}
\newcommand{\rjup}{{R$_\textrm{Jup}$}}
\newcommand{\LA}{{LHS\,6343\,A}}
\newcommand{\LB}{{LHS\,6343\,B}}
\newcommand{\LC}{{LHS\,6343\,C}}
\newcommand{\LHS}{{LHS\,6343}}
\newcommand{\ira}{{3.6$\mu$m}}
\newcommand{\irb}{{4.5$\mu$m}}

\begin{document}
\title{Benchmark Transiting Brown Dwarf LHS\,6343\,C: \textit{SPITZER} Secondary Eclipse Observations Yield 
Brightness Temperature and mid-T Spectral Class}

\author{Benjamin T. Montet\altaffilmark{1,2}, 
John Asher Johnson\altaffilmark{2}, 
Jonathan J. Fortney\altaffilmark{3}, Jean-Michel Desert\altaffilmark{4}}

\email{btm@astro.caltech.edu}

\altaffiltext{1}{Cahill Center for Astronomy and Astrophysics, California Institute of Technology, 1200 E. California Blvd., MC 249-17, Pasadena, CA 91106, USA}
\altaffiltext{2}{Harvard-Smithsonian Center for Astrophysics, 60 Garden
Street, Cambridge, MA 02138, USA}
\altaffiltext{3}{Department of Astronomy and Astrophysics, University of California,
Santa Cruz, CA 95064, USA}
\altaffiltext{4}{Anton Pannenkoek Astronomical Institute, University of Amsterdam, 1090 GE Amsterdam, The Netherlands}

\date{\today, \currenttime}

\begin{abstract}
There are no field brown dwarf analogs with measured masses, radii, and luminosities,
 precluding our ability to connect the population of transiting brown dwarfs with 
measurable masses and radii and field brown dwarfs with measurable luminosities and
atmospheric properties.
\LC, a weakly-irradiated brown dwarf transiting one member of an M+M binary in the \itk\ field, provides
the first opportunity to probe the atmosphere of a non-inflated brown dwarf with a
measured mass and radius. 
Here, we analyze four \spitz\ observations of secondary eclipses of \LC\ behind \LA.
Jointly fitting the eclipses with a Gaussian process noise model of the instrumental 
systematics, we measure eclipse depths of $1.06 \pm 0.21$ ppt at \ira\ and $2.09 \pm 0.08$ ppt
at \irb, corresponding to brightness temperatures of $1026 \pm 57$ K and $1249 \pm 36$ K, respectively.
We then apply brown dwarf evolutionary models to infer a bolometric luminosity 
$\log(L_\star /L_\odot) = -5.16  \pm   0.04$.
Given the known physical properties of the brown dwarf and the two M dwarfs in the \LHS\ 
system, these depths are consistent with models of a 1100 K T dwarf at an age of
5 Gyr and empirical observations of field T5-6 dwarfs with temperatures of $1070 \pm 130$ K.
We investigate the possibility that the orbit of \LC\ has been altered by the Kozai-Lidov
mechanism and propose additional astrometric or Rossiter-McLaughlin measurements of
the system to probe the dynamical history of the system.
\end{abstract}


\maketitle

\section{Introduction}
\label{sec:intro}

There are only
eleven brown dwarfs with measured 
masses and radii \citep[hereafter M15, and references therein]{Montet15a}.
These objects serve as useful benchmark stars to compare theoretical 
predictions of physical parameters for the thousands of known brown dwarfs with
measured luminosities, colors, or other atmospheric parameters 
\citep{Faherty13, Mace13, Helling14}.
Such comparisons are not currently possible as the only brown dwarfs with measured
masses and radii and inferred atmospheric parameters are larger than field objects 
due to youth or irradiation
and therefore not representative of their old, isolated counterparts 
\citep{Stassun06, Siverd12}.

Recently, M15 announced refined physical properties of the 
brown dwarf \LC\ \citep{Johnson11a}, measuring a mass of $62.1 \pm 1.2$ \mjup{} and a radius of 
$0.783 \pm 0.011$ \rjup. These authors also detected a secondary eclipse in the
\itk\ dataset with a depth of $25 \pm 7$ ppm.
This $3.6 \sigma$ detection is insufficient for atmospheric characterization,
but it allows for the possibility of observations at other 
wavelengths to probe the temperature, age, and atmospheric properties of the 
brown dwarf.
\LC\ presents the first opportunity to robustly measure the
atmospheric properties of an old, non-inflated brown dwarf with a known mass
and radius, enabling a key connection between the field and transiting 
brown dwarf populations.

\spitz\ \citep{Werner04} enables observations of the secondary eclipse of
\LC\ behind \LA, providing an opportunity to measure the emitted 
near-IR radiation from the brown dwarf. 
Given the low level of irradiation from the host star, \LC\ should behave 
like a field brown dwarf for which direct mass and radius measurements 
are generally unobtainable (\textsection 5.1)

In this paper, we present detections of the secondary eclipse of \LC\ in both
 \spitz\ IRAC bandpasses.
We measure the eclipse depths by jointly fitting a Gaussian process (GP) model
to the instrumental systematics and a physical model of the astrophysical
signal.
We use these data to infer a temperature and age of the system through 
theoretical models of brown dwarf evolution, making \LC\ the first 
non-inflated brown dwarf with a known mass, radius, and direct measurement
of its atmospheric properties.

\section{Data Collection and Analysis}

We collected data during four separate eclipses with \spitz, two 
each in the 3.6 and 4.5 $\mu$m IRAC bands \citep{Fazio04}. 
These data were collected on 2014 July 06, July 19, September 21, and October 16
as a part of \spitz\ Cycle 10 program 10122 (PI Montet).
Data in both bandpasses were collected in subarray mode with 2.0 second exposures.
In all observations, a 30-minute peak-up preceded the science observations
to place the star on the detector ``sweet-spot'' to minimize pixel-phase effects
\citep[e.g.][]{Ballard10}.
Each set of science observations contains a total of 8768 frames spread over
4.9 hours approximately centered on the time of eclipse.
For computational feasibility, we binned the observations by a factor of
eight, giving a cadence of $\approx$16 seconds per binned data point,
shorter than any astrophysical quantity of interest.

We measure the observed flux in each binned frame by performing aperture photometry,
repeating this procedure 11 times with circular apertures between 1.6 and 3.5 pixels.
By fitting a two-dimensional
Gaussian to the 5x5 region of the detector directly surrounding the brightest pixel,
we measure the position of the star on the detector in each frame \citep{Agol10}.
We find a scatter of $\sim$0.1 pixels during each observation.
A background estimate is calculated by fitting a Gaussian
to the histogram of flux values obtained over each full frame.

\subsection{Noise Model}

The \spitz\ light curves are dominated by instrumental systematics largely caused by 
intrapixel variability in the sensitivity of the InSb detector \citep{Charbonneau05, Knutson08}.
To account for these systematics, we fit an instrumental model 
simultaneously with our secondary eclipse model.
Our instrumental model is the GP model of \citet{Evans15},
who employ a covariance kernel which is a function of the centroid $xy$ coordinates of the
star and the time $t$ of the observation. 
For any two points $i$ and $j$, their covariance is defined such that
\begin{equation}
K_{ij} = k_{xy} + k_t,
\end{equation}
where
\begin{equation}
k_{xy} = A_{xy}^2 \exp\bigg[-\bigg(\frac{x_i - x_j}{L_x}\bigg)^2 - \bigg(\frac{y_i - y_j}{L_y}\bigg)^2 \bigg]
\end{equation}
and
\begin{equation}
k_t = A_t^2 \bigg[1 + \frac{t_i - t_j}{L_t} \sqrt{3} \bigg] \exp \bigg[ - \bigg(\frac{t_i - t_j}{L_t}\bigg) \sqrt{3}\bigg].
\end{equation}
Here, $x_i$ and $y_i$ are the centroid positions of the star during the $i$th observation, taken at time $t_i$. 
$A_{xy}$ and $A_t$ define the magnitude of the correlation between data points and 
$L_x$, $L_y$, and $L_t$ define the length scales of said correlation. 
A larger value of $K_{ij}$, when the temporal or spatial separation between two points is small
relative to $L_t$, $L_x$, or $L_y$, implies a stronger correlation.

Our noise model then has 19 free parameters. As each observation falls on a different region
of the detector, $A_{xy}$, $A_t$, $L_x$, and $L_y$ are not shared between observations. 
$L_t$ is shared between observations. 
We also fit for two white noise parameters, one for each bandpass, added in quadrature to our covariance kernels.

\subsection{Physical Model}

\begin{figure*}[htbp!]
\centerline{\includegraphics[width=0.75\textwidth]{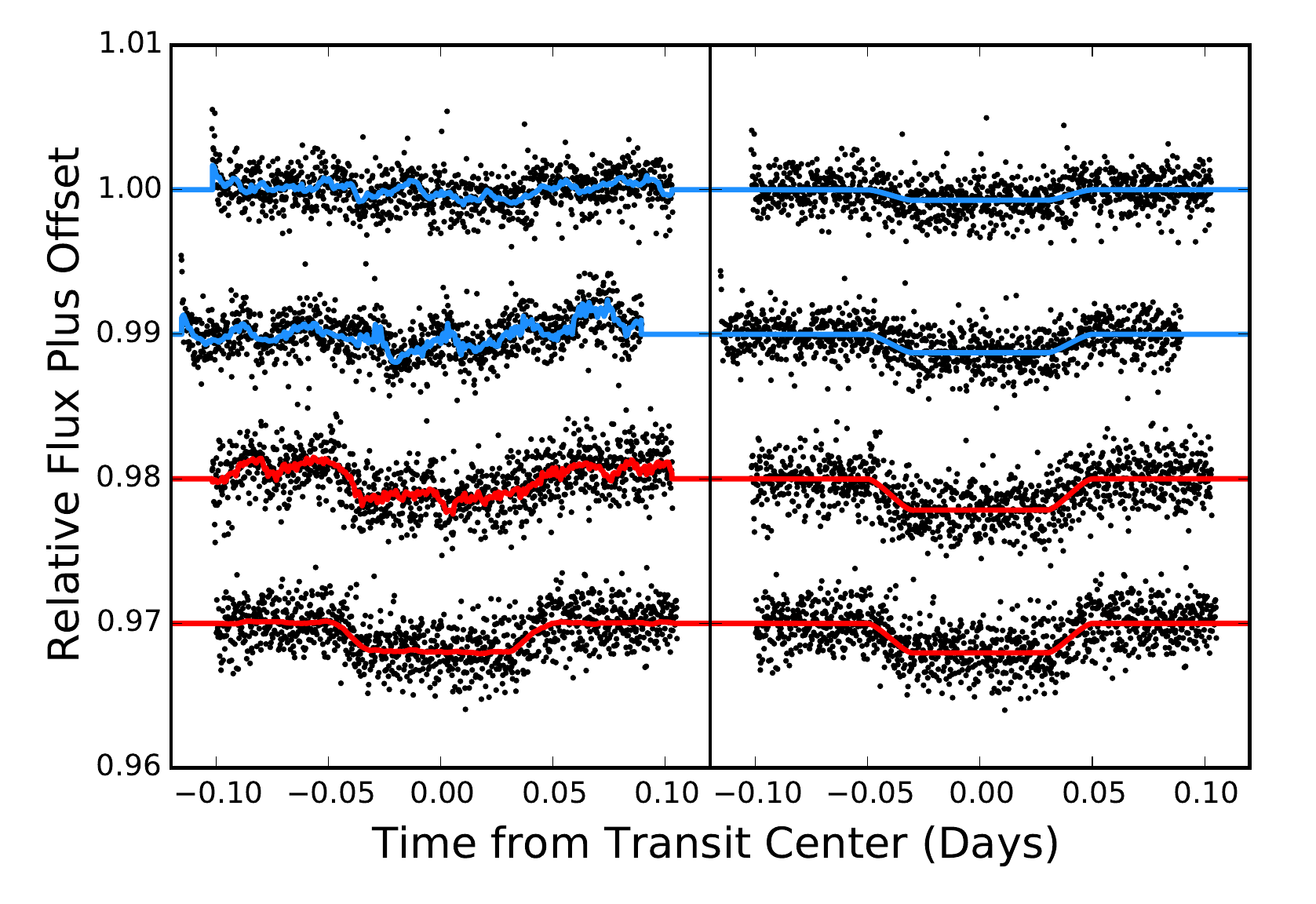}}
\caption{(Left) Observed secondary eclipses of \LC. The solid line represents the maximum likelihood
joint fit of the instrumental and astrophysical models. The observations are arranged chronologically from
top to bottom. The top two, in blue, are eclipses in the IRAC 1 \ira\ bandpass. The bottom two,
in red, are taken in the IRAC 2 \irb\ bandpass. (Right) The same eclipses, with the maximum likelihood
instrumental model divided out for illustration.}
\label{fig:eclipses}
\end{figure*}

Simultaneously we fit a physical model of the secondary eclipses of \LC\ behind \LA. 
We use the transit model of \citet{Mandel02} with no limb darkening, as the primary star
is not being occulted:
the observed flux should be unchanging between second and third contact. 
We fit for four separate eclipse depths, allowing for the possibility of variability 
similar to that observed in \spitz\ surveys of field brown dwarfs \citep{Buenzli12, Metchev15}.
We also fit the orbital period, radius ratio between \LA\ and \LC, time of transit,
eccentricity vectors $\sqrt{e} \cos \omega$ and $\sqrt{e} \sin \omega$, reduced semimajor
axis $a/R_\star$, and impact parameter.
For each of these, we apply a prior following the results of the simultaneous RV and transit 
fit of M15. 
With 11 parameters defining the astrophysical model, we have 30 parameters total.
Our model is shown in Figure \ref{fig:eclipses}.

\subsection{Parameter Estimation}
We first calculate a maximum likelihood solution for each eclipse with each of our 
eleven apertures. We then choose the single aperture which maximizes our likelihood
function and restrict ourselves to that aperture. 
For the first \ira\ eclipse and both \irb\ eclipses, we find the likelihood
function is maximized with a 2.0 pixel aperture;
for the other \ira\ eclipse, we use a 2.3 pixel aperture.
In all cases, these apertures include both M dwarfs in the system.
To compute the covariance matrix and likelihood function for each model,
we use \texttt{george}\footnote{http://dan.iel.fm/george}, 
an implementation of the hierarchically off-diagonal low-rank matrix solver of
\citet{Ambikasaran15}.

To infer the eclipse depths, we then explore the parameter space 
using \texttt{emcee} \citep{Foreman-Mackey12},
an implementation of the affine-invariant ensemble sampler of \citet{Goodman10}.
We initialize 200 walkers clustered around the maximum likelihood
values for each eclipse. 
We then allow these walkers to evolve for 1,500 steps, limiting 
each noise parameter to values within a factor of $e^{10}$ of the maximum 
likelihood value.
We remove the first 600 steps as burn-in and
verify our system has converged through the test of \citet{Geweke92} and 
visual inspection.

\section{Results}

\begin{figure}[htbp!]
\centerline{\includegraphics[width=0.45\textwidth]{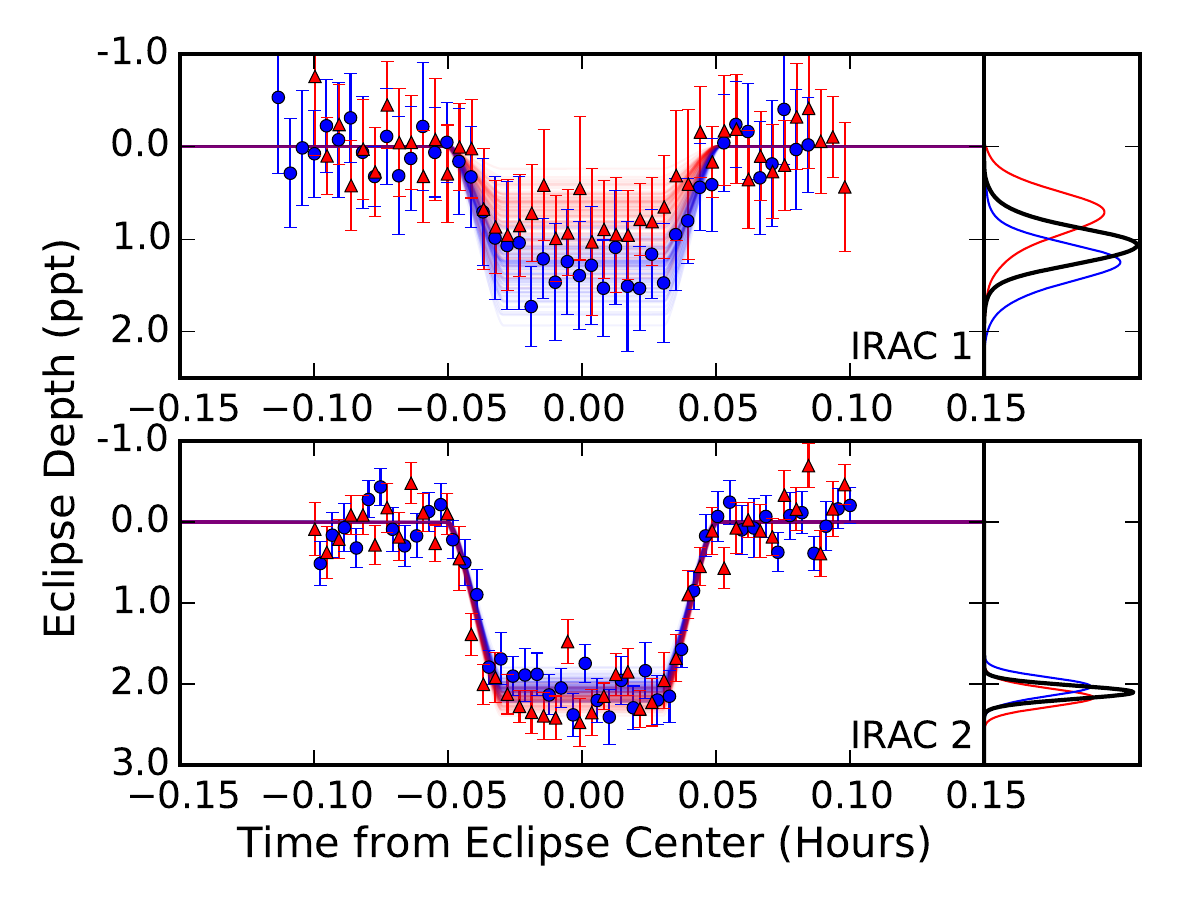}}
\caption{(Left) Observed secondary eclipses in each bandpass, with different transits in
each bandpass labeled with red triangles and blue circles. A representative instrumental
model has been removed for clarity. Red and blue lines represent 
draws from the transit model posterior distributions.
(Right) Marginalized posterior distributions of the eclipse depth for each individual
transit (red, blue) and combined (black). The observed eclipse depths are
consistent at $1.4 \sigma$ in the \ira\ IRAC 1 bandpass and 
$0.8\sigma$ in the \irb\ IRAC 2 bandpass. We find depths of
$1.06 \pm 0.21$ ppt at \ira\ and $2.09 \pm 0.08$ ppt at \irb. }
\label{fig:depths}
\end{figure}

Our results are shown in Table 1. 
We find less correlated noise in the \irb\ bandpass, in line with
previous \spitz\ analyses \citep{Hora08}.
We do not find significant evidence for variability between eclipses.
In the \ira\ bandpass the two depths are consistent at $1.4 \sigma$;
at \irb, $0.8\sigma$.
We consider these observations to represent the system in similar states
and combine the likelihoods on the eclipse depth through a kernel density
estimation of each individual depth.
From this, we measure an eclipse depth of $1.06 \pm 0.21$ parts per thousand (ppt) at \ira\ and $2.09 \pm 0.08$ ppt
at \irb, as shown in Figure \ref{fig:depths}.
We also calculate brightness temperatures for each bandpass using the BT-Settl model spectra of \citet{Allard12}
to infer the expected blackbody flux from the brown dwarf, finding $T_b = 1026 \pm 57$ K at \ira\ and $T_b = 
1249 \pm 36$ K at \irb.

\begin{deluxetable*}{lccc}
\tablecaption{Parameters for AB}
\footnotesize
\tablewidth{0pt}
\tablehead{
  \colhead{Parameter} & 
  \colhead{Median}     &
  \colhead{} &
  \colhead{Uncertainty}  \\
  \colhead{} & 
  \colhead{} &
  \colhead{} &
  \colhead{($1\sigma$)}      
}
\startdata
\textit{IRAC 1 Parameters}\\
Transit Depth, 2014 July 06 (ppt) & 0.74 & $\pm$ & 0.27 \\
Transit Depth, 2014 July 19 (ppt) & 1.26 & $\pm$ & 0.24 \\
Transit Depth, Combined (ppt) & 1.06 & $\pm$ & 0.21 \\
$M_A$ (Vega)\tablenotemark{1} & 6.56  & $\pm$ & 0.08 \\
$M_B$ (Vega)\tablenotemark{1} & 6.97 & $\pm$ & 0.10 \\
$M_C$ (Vega)& 13.43 & $\pm$ & 0.23 \\
$T_b$ (K) & 1026 & $\pm$ & 57 \\
\\
\textit{IRAC 2 Parameters}\\
Transit Depth, 2014 September 21 (ppt) & 2.16 &$\pm$  & 0.12 \\
Transit Depth, 2014 October 16 (ppt) & 2.03 & $\pm$ & 0.12 \\
Transit Depth, Combined (ppt) & 2.09 & $\pm$ & 0.08 \\
$M_A$ (Vega)\tablenotemark{1} & 6.45  & $\pm$ & 0.07 \\
$M_B$ (Vega)\tablenotemark{1} & 6.86 & $\pm$ & 0.09 \\
$M_C$ (Vega) & 12.58 & $\pm$ & 0.07 \\
$T_b$ (K) & 1249 & $\pm$ & 36 \\
\\
\textit{System Parameters} \\
Time of Secondary Eclipse (BJD - 2400000) & 56845.401 & $\pm$ & 0.001 \\
Orbital Period (days)\tablenotemark{2} &  12.7137941 & $\pm$ & 0.0000002 \\
Eccentricity Vector $e \cos \omega$ & 0.0229 & $\pm$ & 0.0001 \\
Star C Surface Gravity (m s$^{-2}$)\tablenotemark{2} & 2630 & $\pm$ & 50  \\
Star C Luminosity ($\log(L_\star /L_\odot$))\tablenotemark{3}  & -5.16 & $\pm$  & $0.04$  \\
Star C Temperature\tablenotemark{3} (K) & 1130 & $\pm$ & 50 \\
Star C Age (Gyr)\tablenotemark{3} & 5 & $\pm$ & 1

\enddata
\tablenotetext{1}{Inferred through $VRJHK$ photometry and the Dartmouth models of \citet{Dotter08}  }
\tablenotetext{2}{From M15}
\tablenotetext{3}{Dependent on the BT-Settl evolutionary models of \citet{Allard12}}
\label{tab:results}
\end{deluxetable*}

To test the robustness of our GP model, we calculate the maximum
likelihood solutions with two different instrumental models.
Following \citet{Knutson08}, we fit a second-order polynomial
to the inferred centroid positions of the star to decorrelate the telescope motion
from the astrophysical signal. 
We also apply the pixel-level decorrelation method of \citet{Deming15}, which
decorates the observed fluxes against the pixel counts inside a subarray centered on the
PSF of the star.
In both cases, we find no statistical difference on the inferred eclipse depths.

\section{Temperature and Age of \LC}

Given the \spitz\ eclipse depths and the known mass and radius of \LC, 
we can infer the temperature of \LC\ and the age of the system. 
The eclipse depths only provide a ratio between the flux
from the brown dwarf and the two M dwarfs:
\begin{equation}
\delta = \frac{F_C}{F_A + F_B + F_C}.
\end{equation}
We have no direct measurement of the brightness of the two M dwarfs in the
IRAC bandpasses so we must infer them.
M15 use the Dartmouth stellar evolutionary models of \citet{Dotter08}
to infer a mass and radius for each star given available $VRJHK$ photometry.
Here, we use the posterior distributions on the stellar masses and the Dartmouth models to predict
the absolute magnitudes of the stars at 3.6 and \irb (Table 1).
This technique also reproduces the expected brightness of the M dwarfs to within the
photometric uncertainties in all bandpasses where we do have data.
We then use these predictions and the observed eclipse depths to 
calculate the absolute magnitude of \LC\ in 
both IRAC bandpasses: we determine $M_{C, 3.6} = 13.43 \pm 0.23$ and 
$M_{C, 4.5} = 12.58 \pm 0.07$ so that $[3.6 - 4.5] = 0.85 \pm 0.24$.
We repeat this procedure with the resolved flux measurements and the BT-Settl
evolutionary models of \citet{Allard12}, finding no difference in the 
extrapolated IRAC absolute magnitudes of the M dwarfs at the $1\sigma$ level.

Brown dwarf evolutionary models can be used to determine a temperature
and age of \LC.
We investigate the predictions of several models.

\begin{figure}[htbp!]
\centerline{\includegraphics[width=0.45\textwidth]{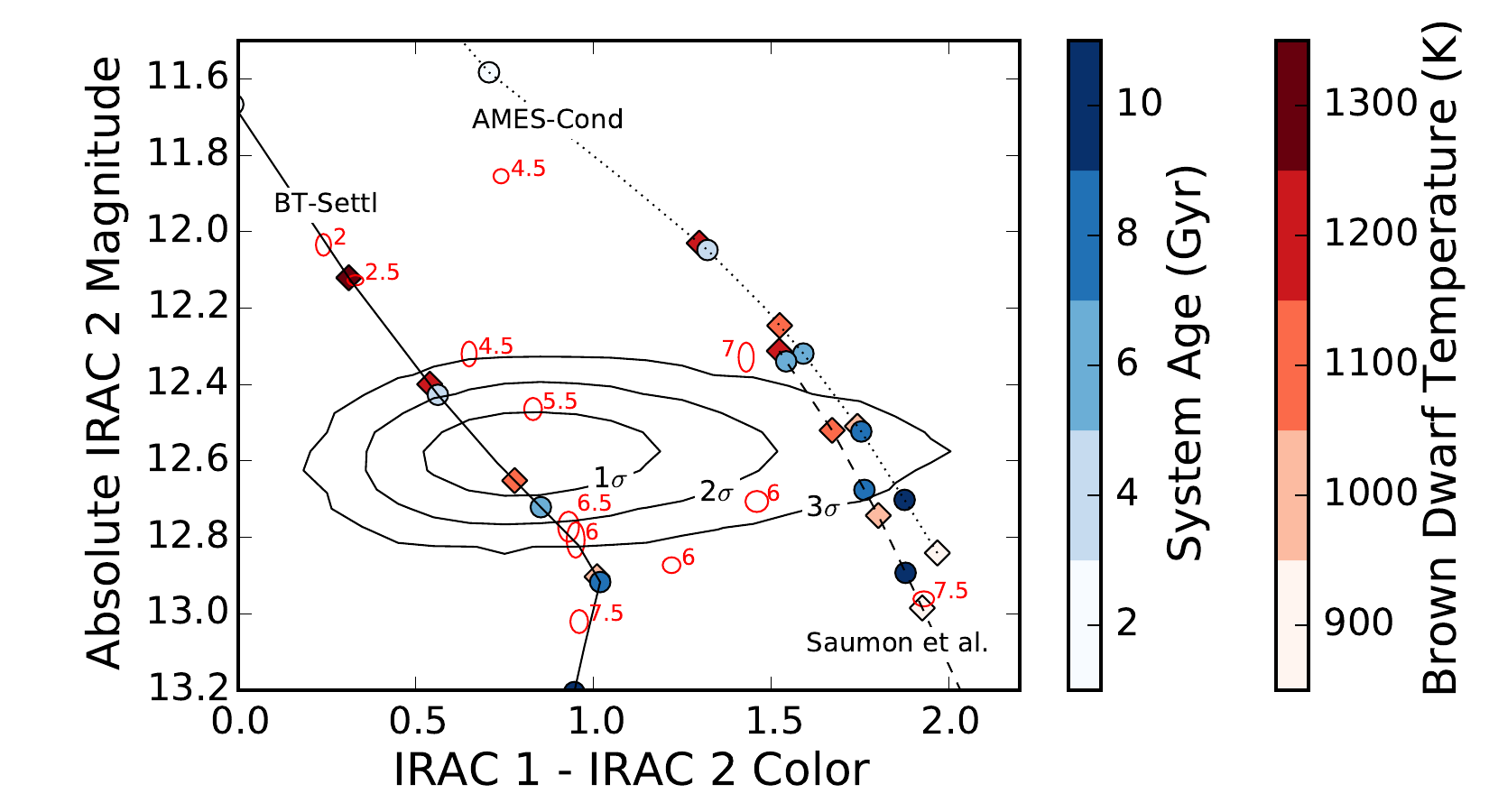}}
\caption{Color-magnitude diagram showing the absolute magnitude in the IRAC 2 \irb\ bandpass
against the IRAC 1 - IRAC 2 color. Contours represent the allowed parameter space in which \LC\ could 
reside. The labeled lines represent the theoretical evolutionary tracks of a brown dwarf with the mass of 
\LC\ from (left to right) the BT-Settl, \citet{Saumon12}, and AMES-Cond models. Dots correspond to model predictions
at (white to dark blue) 2, 4, 6, 8, and 10 Gyr; diamonds correspond to model predictions for temperatures of
(white to dark red) 900, 1000, 1100, 1200, and 1300 K.The BT-Settl model provides a good fit at $5 \pm 1$ Gyr; the 
AMES-Cond and Saumon models fit the data at lower significance at $8 \pm 1$ Gyr. 
Red ellipses represent field brown dwarfs from \citet{Dupuy12} and \citet{Filippazzo15}; labels represent the spectral subtype inside the T class.  }
\label{fig:CMD}
\end{figure}

The BT-Settl models provide the best fit to the available data. We use the isochrones
calculated for the CIFIST 2011 abundances and opacities \citep{Caffau10, Allard12}, the most recent for which 
magnitudes have been tabulated at these masses and ages.
With this model grid, we infer a brown dwarf with $t = 5 \pm 1$ Gyr, $T = 1130 \pm 50$ K,
and $\log(L_\star /L_\odot = -5.16  \pm   0.04$ by evaluating the likelihood of the model fit to our calculated absolute magnitudes
in each bandpass and marginalizing over all other parameters. 
This strategy provides an estimate of the statistical error, but not the systematic error caused by uncertainty
or errors in the models.
We note that of field brown dwarfs with measured temperatures
and colors, this model set predicts the correct temperatures with a scatter of $\sim 50$K, consistent with 
the published uncertainties in temperature.

The AMES-Cond models of \citet{Allard01} provide a fit to the \spitz\ photometry,
mass, and radius of \LC\ such that $t = 8 \pm 1$~Gyr and $T = 1000 \pm 50$~K.
However, this model grid underpredicts the \ira\ luminosity, leading to an overestimation
of the [3.6 - 4.5] color at all ages (Figure \ref{fig:CMD}).
The AMES-Dusty models, meanwhile, do not provide a good fit, overpredicting the luminosity
even if the system were the age of the universe, as is common with brown dwarf models
\citep{Rice10, Dupuy15}.

The isochrones of \citet{Saumon08} combined with synthetic photometry from \citet{Saumon12} 
predict IRAC photometry as a function of temperature and system age.
These models provide a slightly better fit to the data than the AMES-Cond
models for an $1100 \pm 50$ K brown dwarf, but still overpredict the [3.6 - 4.5] color.
Their hybrid models, meant to model the L/T transition, suggest an older brown dwarf with an age
of $8 \pm 1$ Gyr.
Their cloudy L dwarf models do not provide a good fit at any age.
Given the inability of the cloudy models to explain the observations, as well as the consistency 
between models in predicting temperatures below the L/T transition
\citep{Burgasser02, Golimowski04}, we confirm \LC\ as a T dwarf.

Objects near the L/T transition with temperatures 1000-1400 K
are particularly challenging for brown dwarf evolutionary models.
The uncertainties in all models are dominated by systematics, so we
cannot develop one statistical posterior on the 
temperature or age. 
We note the BT-Settl models provide the best fit to these data and to the population of similar mid-T dwarfs in color-magnitude
space.
This system compares favorably to other known T5-6 dwarfs \citep[Figure 3]{Dupuy12}. Of the field brown dwarfs with measured luminosities and temperatures, it is consistent
with being between T4.5 dwarf 2MASS\,0000+2554 ($1227 \pm 95$ K) and T6 dwarfs 2MASS\,0243-2453 ($973 \pm 83$ K) and 2MASS\,1346-0031 \citep[$1011 \pm 86$ K,][]{Filippazzo15} in its evolution.
This age measurement, while model dependent, is the first measurement of the age of the system: previously, 
\citet{Johnson11a} were able to only place a lower limit of 1-2 Gyr on the system age.

\section{Discussion}

\subsection{Irradiation from \LA}

We ignore irradiation from \LA.
Given the (Dartmouth model-dependent) temperature of the host of $3430 \pm 20$ K 
and semi major axis $a/R_\star = 46.0 \pm 0.4$, the equilibrium temperature
of the brown dwarf is $T_{eq} = 365\pm 3$ K, assuming a Bond albedo of 0.07, expected for 
a massive brown dwarf around an M2V dwarf \citep{Marley99}.
Therefore, the emitted flux as a result of the absorption and reemission of stellar 
radiation from \LA\ is $\approx 1\%$ the total flux. 
While irradiation may affect the thermal profile of the brown dwarf, it should
be negligible considering the $\approx 0.1$~mag uncertainties on the brown dwarf's 
magnitude.

Moreover, given the advanced age of the system, we expect high energy 
irradiation from the host star to be negligible. \citet{West08} find a rapid decay
in M dwarf magnetic activity over stellar age; \citet{Shkolnik14}
find the same to be true for UV emission, with a steep drop in UV emission at ages
above $1$ Gyr.
\citet{Stelzer13} study nearby M dwarfs to find X-ray emission decays even more quickly
for M dwarfs than UV emission, with a difference of three orders of magnitude between 
young M dwarfs in TW Hydra and old, field M dwarfs.
Any high energy radiation that may have once influenced the atmosphere of \LC\ has 
been at a low level for billions of years, allowing the brown dwarf to 
achieve an equilibrium representative of field brown dwarfs.

\subsection{Metallicity of \LC}

M15 infer a metallicity for the two M dwarfs in the system 
$[\textrm{a/H}] = 0.02 \pm 0.19$.
If the brown dwarf formed through core accretion, it may be expected to have
a higher metallicity than its host stars \citep{Pollack86, Podolak88},
as is the case for the planet orbiting GJ\,504 \citep{Skemer15}. 
Because of the low mass of the host star and likely low mass of its protoplanetary disk
\citep{Andrews13}, it is considerably more likely this brown dwarf formed like a binary
star system so that the metallicity of \LC\ is likely not significantly different from its
host star \citep{Desidera04}.
Additional observations that infer a spectrum of \LC\ can provide 
tests of theoretical brown dwarf spectra given the known metallicity of the system.
These tests are especially important for mid/late T dwarfs, where metallicity effects
can affect near-IR colors by as much as 0.3 dex \citep{Burningham13}.

\subsection{Dynamical History of \LHS}

The secondary eclipses are centered at phase $0.5146 \pm 0.0001$, corresponding
to times of transit $0.185 \pm 0.001$ days after half-phase between
successive primary transits, or an eccentricity vector 
$e \cos{\omega} = 0.0229 \pm 0.0001$.
This value is consistent with that inferred from RV observations and 
\itk\ photometry ($0.0228 \pm 0.0008$, M15)

The eccentricity in the \LA-C subsystem may be primordial or the result
of dynamical perturbations from star B. 
\LB\ is presently at a sky-projected separation of $\sim$20 AU from the A-C subsystem.
Depending on the orbit of \LB, the system may be susceptible to Kozai-Lidov oscillations
\citep{Kozai62, Lidov62}.
Kozai-Lidov cycles would lead to oscillations in the orbital inclination and eccentricity 
of the A-C subsystem on a timescale
\begin{equation}
\tau \approx P_C \frac{M_{AC}}{M_B} \bigg(\frac{a_{AC-B}}{a_{AC}}\bigg)^3 (1 - e_{AC-B})^{3/2},
\end{equation}
where $P_C$ is the orbital period of the brown dwarf, $M_{AC}$ the A-C subsystem mass,
$M_B$ the perturber mass, $a_{AC-B}$ and $e_{AC-B}$ the orbital semimajor axis and 
eccentricity of star B around the AC subsystem, and $a_{AC}$ the orbital semimajor axis
of C around A.
The two M dwarfs have similar masses. The semimajor axis $a_{AC} = 0.08$ AU is known,
but we only know the instantaneous sky-projected separation between AC and B is $\approx $20 AU.
Taking this value as a proxy for the true semimajor axis, we find
$\tau \sim 10^6 (1-e_{AC-B})^{3/2}$ years. Even for significantly larger orbits of star B and 
high eccentricities, the timescales for Kozai-Lidov cycles would be shorter
than the $\sim 10^{10}$ year age of the system, suggesting the system may be susceptible
to Kozai-Lidov oscillations given appropriate initial conditions.

The current orbit can provide clues about the dynamical history of this system. 
Measurement of an inclined orbit of \LB\ through astrometric 
monitoring could provide evidence for Kozai-Lidov cycles, as would a 
misalignment between the spin axis of \LA\ and the orbit of \LC.
While close binaries are not always neatly aligned \citep{Albrecht14},
they often are, especially for low-mass binaries \citep{Harding13, Triaud13}.

\acknowledgements
We thank the referee, Adam Burgasser, for his thorough referee report which considerably improved the
quality of this paper. We thank Drake Deming for providing an early draft of his 2015 paper and a 
version of the underlying code. 
We thank Sarah Ballard and Dan Foreman-Mackey for conversations about
\spitz\ data analysis and Mark Marley and Jackie Faherty
for very helpful comments on an early draft of this paper which significantly improved
its quality.

This work is based on observations made with the \spitz\ Space Telescope, 
which is operated by the Jet Propulsion Laboratory, California Institute of Technology 
under a contract with NASA. 
Support for this work was provided by NASA through an award issued by JPL/Caltech.

B.T.M. is supported by the National Science Foundation Graduate Research Fellowship under Grant No. DGE-€1144469. 
J.A.J. is supported by generous grants from the David and Lucile Packard Foundation and the Alfred P. Sloan Foundation.

{\it Facilities:} \facility{\spitz\ (IRAC)}


\begin{thebibliography}{}
\providecommand\natexlab[1]{#1}
\providecommand\JournalTitle[1]{#1}

\bibitem[{{Agol} {et~al.}(2010){Agol}, {Cowan}, {Knutson}, {Deming}, {Steffen},
  {Henry}, \& {Charbonneau}}]{Agol10}
{Agol}, E., {Cowan}, N.~B., {Knutson}, H.~A., {et~al.} 2010,
  \href{http://dx.doi.org/10.1088/0004-637X/721/2/1861}{\JournalTitle{\apj},
  721, 1861}

\bibitem[{{Albrecht} {et~al.}(2014){Albrecht}, {Winn}, {Torres}, {Fabrycky},
  {Setiawan}, {Gillon}, {Jehin}, {Triaud}, {Queloz}, {Snellen}, \&
  {Eggleton}}]{Albrecht14}
{Albrecht}, S., {Winn}, J.~N., {Torres}, G., {et~al.} 2014,
  \href{http://dx.doi.org/10.1088/0004-637X/785/2/83}{\JournalTitle{\apj}, 785,
  83}

\bibitem[{{Allard} {et~al.}(2001){Allard}, {Hauschildt}, {Alexander},
  {Tamanai}, \& {Schweitzer}}]{Allard01}
{Allard}, F., {Hauschildt}, P.~H., {Alexander}, D.~R., {Tamanai}, A., \&
  {Schweitzer}, A. 2001,
  \href{http://dx.doi.org/10.1086/321547}{\JournalTitle{\apj}, 556, 357}

\bibitem[{{Allard} {et~al.}(2012){Allard}, {Homeier}, \& {Freytag}}]{Allard12}
{Allard}, F., {Homeier}, D., \& {Freytag}, B. 2012,
  \href{http://dx.doi.org/10.1098/rsta.2011.0269}{\JournalTitle{Philosophical
  Transactions of the Royal Society of London Series A}, 370, 2765}

\bibitem[{{Ambikasaran} {et~al.}(2014){Ambikasaran}, {Foreman-Mackey},
  {Greengard}, {Hogg}, \& {O'Neil}}]{Ambikasaran15}
{Ambikasaran}, S., {Foreman-Mackey}, D., {Greengard}, L., {Hogg}, D.~W., \&
  {O'Neil}, M. 2014, \JournalTitle{ArXiv e-prints},
  \href{http://arxiv.org/abs/1403.6015}{{\sffamily arXiv:1403.6015 [math.NA]}}

\bibitem[{{Andrews} {et~al.}(2013){Andrews}, {Rosenfeld}, {Kraus}, \&
  {Wilner}}]{Andrews13}
{Andrews}, S.~M., {Rosenfeld}, K.~A., {Kraus}, A.~L., \& {Wilner}, D.~J. 2013,
  \href{http://dx.doi.org/10.1088/0004-637X/771/2/129}{\JournalTitle{\apj},
  771, 129}

\bibitem[{{Ballard} {et~al.}(2010){Ballard}, {Charbonneau}, {Deming},
  {Knutson}, {Christiansen}, {Holman}, {Fabrycky}, {Seager}, \&
  {A'Hearn}}]{Ballard10}
{Ballard}, S., {Charbonneau}, D., {Deming}, D., {et~al.} 2010,
  \href{http://dx.doi.org/10.1086/657159}{\JournalTitle{\pasp}, 122, 1341}

\bibitem[{{Buenzli} {et~al.}(2012){Buenzli}, {Apai}, {Morley}, {Flateau},
  {Showman}, {Burrows}, {Marley}, {Lewis}, \& {Reid}}]{Buenzli12}
{Buenzli}, E., {Apai}, D., {Morley}, C.~V., {et~al.} 2012,
  \href{http://dx.doi.org/10.1088/2041-8205/760/2/L31}{\JournalTitle{\apjl},
  760, L31}

\bibitem[{{Burgasser} {et~al.}(2002){Burgasser}, {Kirkpatrick}, {Brown},
  {Reid}, {Burrows}, {Liebert}, {Matthews}, {Gizis}, {Dahn}, {Monet}, {Cutri},
  \& {Skrutskie}}]{Burgasser02}
{Burgasser}, A.~J., {Kirkpatrick}, J.~D., {Brown}, M.~E., {et~al.} 2002,
  \href{http://dx.doi.org/10.1086/324033}{\JournalTitle{\apj}, 564, 421}

\bibitem[{{Burningham} {et~al.}(2013){Burningham}, {Cardoso}, {Smith},
  {Leggett}, {Smart}, {Mann}, {Dhital}, {Lucas}, {Tinney}, {Pinfield}, {Zhang},
  {Morley}, {Saumon}, {Aller}, {Littlefair}, {Homeier}, {Lodieu}, {Deacon},
  {Marley}, {van Spaandonk}, {Baker}, {Allard}, {Andrei}, {Canty}, {Clarke},
  {Day-Jones}, {Dupuy}, {Fortney}, {Gomes}, {Ishii}, {Jones}, {Liu},
  {Magazz{\'u}}, {Marocco}, {Murray}, {Rojas-Ayala}, \&
  {Tamura}}]{Burningham13}
{Burningham}, B., {Cardoso}, C.~V., {Smith}, L., {et~al.} 2013,
  \href{http://dx.doi.org/10.1093/mnras/stt740}{\JournalTitle{\mnras}, 433,
  457}

\bibitem[{{Caffau} {et~al.}(2010){Caffau}, {Ludwig}, {Bonifacio}, {Faraggiana},
  {Steffen}, {Freytag}, {Kamp}, \& {Ayres}}]{Caffau10}
{Caffau}, E., {Ludwig}, H.-G., {Bonifacio}, P., {et~al.} 2010,
  \href{http://dx.doi.org/10.1051/0004-6361/200912227}{\JournalTitle{\aap},
  514, A92}

\bibitem[{{Charbonneau} {et~al.}(2005){Charbonneau}, {Allen}, {Megeath},
  {Torres}, {Alonso}, {Brown}, {Gilliland}, {Latham}, {Mandushev}, {O'Donovan},
  \& {Sozzetti}}]{Charbonneau05}
{Charbonneau}, D., {Allen}, L.~E., {Megeath}, S.~T., {et~al.} 2005,
  \href{http://dx.doi.org/10.1086/429991}{\JournalTitle{\apj}, 626, 523}

\bibitem[{{Deming} {et~al.}(2015){Deming}, {Knutson}, {Kammer}, {Fulton},
  {Ingalls}, {Carey}, {Burrows}, {Fortney}, {Todorov}, {Agol}, {Cowan},
  {Desert}, {Fraine}, {Langton}, {Morley}, \& {Showman}}]{Deming15}
{Deming}, D., {Knutson}, H., {Kammer}, J., {et~al.} 2015,
  \href{http://dx.doi.org/10.1088/0004-637X/805/2/132}{\JournalTitle{\apj},
  805, 132}

\bibitem[{{Desidera} {et~al.}(2004){Desidera}, {Gratton}, {Scuderi}, {Claudi},
  {Cosentino}, {Barbieri}, {Bonanno}, {Carretta}, {Endl}, {Lucatello},
  {Martinez Fiorenzano}, \& {Marzari}}]{Desidera04}
{Desidera}, S., {Gratton}, R.~G., {Scuderi}, S., {et~al.} 2004,
  \href{http://dx.doi.org/10.1051/0004-6361:20041242}{\JournalTitle{\aap}, 420,
  683}

\bibitem[{{Dotter} {et~al.}(2008){Dotter}, {Chaboyer}, {Jevremovi{\'c}},
  {Kostov}, {Baron}, \& {Ferguson}}]{Dotter08}
{Dotter}, A., {Chaboyer}, B., {Jevremovi{\'c}}, D., {et~al.} 2008,
  \href{http://dx.doi.org/10.1086/589654}{\JournalTitle{\apjs}, 178, 89}

\bibitem[{{Dupuy} \& {Liu}(2012)}]{Dupuy12}
{Dupuy}, T.~J., \& {Liu}, M.~C. 2012,
  \href{http://dx.doi.org/10.1088/0067-0049/201/2/19}{\JournalTitle{\apjs},
  201, 19}

\bibitem[{{Dupuy} {et~al.}(2015){Dupuy}, {Liu}, {Leggett}, {Ireland}, {Chiu},
  \& {Golimowski}}]{Dupuy15}
{Dupuy}, T.~J., {Liu}, M.~C., {Leggett}, S.~K., {et~al.} 2015,
  \href{http://dx.doi.org/10.1088/0004-637X/805/1/56}{\JournalTitle{\apj}, 805,
  56}

\bibitem[{{Evans} {et~al.}(2015){Evans}, {Aigrain}, {Gibson}, {Barstow},
  {Amundsen}, {Tremblin}, \& {Mourier}}]{Evans15}
{Evans}, T.~M., {Aigrain}, S., {Gibson}, N., {et~al.} 2015,
  \href{http://dx.doi.org/10.1093/mnras/stv910}{\JournalTitle{\mnras}, 451,
  680}

\bibitem[{{Faherty} {et~al.}(2013){Faherty}, {Rice}, {Cruz}, {Mamajek}, \&
  {N{\'u}{\~n}ez}}]{Faherty13}
{Faherty}, J.~K., {Rice}, E.~L., {Cruz}, K.~L., {Mamajek}, E.~E., \&
  {N{\'u}{\~n}ez}, A. 2013,
  \href{http://dx.doi.org/10.1088/0004-6256/145/1/2}{\JournalTitle{\aj}, 145,
  2}

\bibitem[{{Fazio} {et~al.}(2004){Fazio}, {Hora}, {Allen}, {Ashby}, {Barmby},
  {Deutsch}, {Huang}, {Kleiner}, {Marengo}, {Megeath}, {Melnick}, {Pahre},
  {Patten}, {Polizotti}, {Smith}, {Taylor}, {Wang}, {Willner}, {Hoffmann},
  {Pipher}, {Forrest}, {McMurty}, {McCreight}, {McKelvey}, {McMurray}, {Koch},
  {Moseley}, {Arendt}, {Mentzell}, {Marx}, {Losch}, {Mayman}, {Eichhorn},
  {Krebs}, {Jhabvala}, {Gezari}, {Fixsen}, {Flores}, {Shakoorzadeh}, {Jungo},
  {Hakun}, {Workman}, {Karpati}, {Kichak}, {Whitley}, {Mann}, {Tollestrup},
  {Eisenhardt}, {Stern}, {Gorjian}, {Bhattacharya}, {Carey}, {Nelson},
  {Glaccum}, {Lacy}, {Lowrance}, {Laine}, {Reach}, {Stauffer}, {Surace},
  {Wilson}, {Wright}, {Hoffman}, {Domingo}, \& {Cohen}}]{Fazio04}
{Fazio}, G.~G., {Hora}, J.~L., {Allen}, L.~E., {et~al.} 2004,
  \href{http://dx.doi.org/10.1086/422843}{\JournalTitle{\apjs}, 154, 10}

\bibitem[{{Filippazzo} {et~al.}(2015){Filippazzo}, {Rice}, {Faherty}, {Cruz},
  {Van Gordon}, \& {Looper}}]{Filippazzo15}
{Filippazzo}, J.~C., {Rice}, E.~L., {Faherty}, J., {et~al.} 2015,
  \href{http://dx.doi.org/10.1088/0004-637X/810/2/158}{\JournalTitle{\apj},
  810, 158}

\bibitem[{{Foreman-Mackey} {et~al.}(2013){Foreman-Mackey}, {Hogg}, {Lang}, \&
  {Goodman}}]{Foreman-Mackey12}
{Foreman-Mackey}, D., {Hogg}, D.~W., {Lang}, D., \& {Goodman}, J. 2013,
  \href{http://dx.doi.org/10.1086/670067}{\JournalTitle{\pasp}, 125, 306}

\bibitem[{{Geweke}(1992)}]{Geweke92}
{Geweke}, J. 1992, Bayesian Statistics IV. Oxford: Clarendon Press, ed. J.~M.
  {Bernardo}, 169

\bibitem[{{Golimowski} {et~al.}(2004){Golimowski}, {Henry}, {Krist},
  {Dieterich}, {Ford}, {Illingworth}, {Ardila}, {Clampin}, {Franz},
  {Wasserman}, {Benedict}, {McArthur}, \& {Nelan}}]{Golimowski04}
{Golimowski}, D.~A., {Henry}, T.~J., {Krist}, J.~E., {et~al.} 2004,
  \href{http://dx.doi.org/10.1086/423911}{\JournalTitle{\aj}, 128, 1733}

\bibitem[{Goodman \& Weare(2010)}]{Goodman10}
Goodman, J., \& Weare, J. 2010,
  \href{http://dx.doi.org/10.2140/camcos.2010.5.65}{\JournalTitle{Communications
  in Applied Mathematics and Computational Science}, 5, 65}

\bibitem[{{Harding} {et~al.}(2013){Harding}, {Hallinan}, {Konopacky},
  {Kratter}, {Boyle}, {Butler}, \& {Golden}}]{Harding13}
{Harding}, L.~K., {Hallinan}, G., {Konopacky}, Q.~M., {et~al.} 2013,
  \href{http://dx.doi.org/10.1051/0004-6361/201220865}{\JournalTitle{\aap},
  554, A113}

\bibitem[{{Helling} \& {Casewell}(2014)}]{Helling14}
{Helling}, C., \& {Casewell}, S. 2014,
  \href{http://dx.doi.org/10.1007/s00159-014-0080-0}{\JournalTitle{\aapr}, 22,
  80}

\bibitem[{{Hora} {et~al.}(2008){Hora}, {Carey}, {Surace}, {Marengo},
  {Lowrance}, {Glaccum}, {Lacy}, {Reach}, {Hoffmann}, {Barmby}, {Willner},
  {Fazio}, {Megeath}, {Allen}, {Bhattacharya}, \& {Quijada}}]{Hora08}
{Hora}, J.~L., {Carey}, S., {Surace}, J., {et~al.} 2008,
  \href{http://dx.doi.org/10.1086/593217}{\JournalTitle{\pasp}, 120, 1233}

\bibitem[{{Johnson} {et~al.}(2011){Johnson}, {Apps}, {Gazak}, {Crepp},
  {Crossfield}, {Howard}, {Marcy}, {Morton}, {Chubak}, \&
  {Isaacson}}]{Johnson11a}
{Johnson}, J.~A., {Apps}, K., {Gazak}, J.~Z., {et~al.} 2011,
  \href{http://dx.doi.org/10.1088/0004-637X/730/2/79}{\JournalTitle{\apj}, 730,
  79}

\bibitem[{{Knutson} {et~al.}(2008){Knutson}, {Charbonneau}, {Allen}, {Burrows},
  \& {Megeath}}]{Knutson08}
{Knutson}, H.~A., {Charbonneau}, D., {Allen}, L.~E., {Burrows}, A., \&
  {Megeath}, S.~T. 2008,
  \href{http://dx.doi.org/10.1086/523894}{\JournalTitle{\apj}, 673, 526}

\bibitem[{{Kozai}(1962)}]{Kozai62}
{Kozai}, Y. 1962, \href{http://dx.doi.org/10.1086/108790}{\JournalTitle{\aj},
  67, 591}

\bibitem[{{Lidov}(1962)}]{Lidov62}
{Lidov}, M.~L. 1962,
  \href{http://dx.doi.org/10.1016/0032-0633(62)90129-0}{\JournalTitle{\planss},
  9, 719}

\bibitem[{{Mace} {et~al.}(2013){Mace}, {Kirkpatrick}, {Cushing}, {Gelino},
  {Griffith}, {Skrutskie}, {Marsh}, {Wright}, {Eisenhardt}, {McLean},
  {Thompson}, {Mix}, {Bailey}, {Beichman}, {Bloom}, {Burgasser}, {Fortney},
  {Hinz}, {Knox}, {Lowrance}, {Marley}, {Morley}, {Rodigas}, {Saumon},
  {Sheppard}, \& {Stock}}]{Mace13}
{Mace}, G.~N., {Kirkpatrick}, J.~D., {Cushing}, M.~C., {et~al.} 2013,
  \href{http://dx.doi.org/10.1088/0067-0049/205/1/6}{\JournalTitle{\apjs}, 205,
  6}

\bibitem[{{Mandel} \& {Agol}(2002)}]{Mandel02}
{Mandel}, K., \& {Agol}, E. 2002,
  \href{http://dx.doi.org/10.1086/345520}{\JournalTitle{\apjl}, 580, L171}

\bibitem[{{Marley} {et~al.}(1999){Marley}, {Gelino}, {Stephens}, {Lunine}, \&
  {Freedman}}]{Marley99}
{Marley}, M.~S., {Gelino}, C., {Stephens}, D., {Lunine}, J.~I., \& {Freedman},
  R. 1999, \href{http://dx.doi.org/10.1086/306881}{\JournalTitle{\apj}, 513,
  879}

\bibitem[{{Metchev} {et~al.}(2015){Metchev}, {Heinze}, {Apai}, {Flateau},
  {Radigan}, {Burgasser}, {Marley}, {Artigau}, {Plavchan}, \&
  {Goldman}}]{Metchev15}
{Metchev}, S.~A., {Heinze}, A., {Apai}, D., {et~al.} 2015,
  \href{http://dx.doi.org/10.1088/0004-637X/799/2/154}{\JournalTitle{\apj},
  799, 154}

\bibitem[{{Montet} {et~al.}(2015){Montet}, {Johnson}, {Muirhead}, {Villar},
  {Vassallo}, {Baranec}, {Law}, {Riddle}, {Marcy}, {Howard}, \&
  {Isaacson}}]{Montet15a}
{Montet}, B.~T., {Johnson}, J.~A., {Muirhead}, P.~S., {et~al.} 2015,
  \href{http://dx.doi.org/10.1088/0004-637X/800/2/134}{\JournalTitle{\apj},
  800, 134}

\bibitem[{{Podolak} {et~al.}(1988){Podolak}, {Pollack}, \&
  {Reynolds}}]{Podolak88}
{Podolak}, M., {Pollack}, J.~B., \& {Reynolds}, R.~T. 1988,
  \href{http://dx.doi.org/10.1016/0019-1035(88)90090-5}{\JournalTitle{Icarus},
  73, 163}

\bibitem[{{Pollack} {et~al.}(1986){Pollack}, {Podolak}, {Bodenheimer}, \&
  {Christofferson}}]{Pollack86}
{Pollack}, J.~B., {Podolak}, M., {Bodenheimer}, P., \& {Christofferson}, B.
  1986,
  \href{http://dx.doi.org/10.1016/0019-1035(86)90123-5}{\JournalTitle{Icarus},
  67, 409}

\bibitem[{{Rice} {et~al.}(2010){Rice}, {Barman}, {Mclean}, {Prato}, \&
  {Kirkpatrick}}]{Rice10}
{Rice}, E.~L., {Barman}, T., {Mclean}, I.~S., {Prato}, L., \& {Kirkpatrick},
  J.~D. 2010,
  \href{http://dx.doi.org/10.1088/0067-0049/186/1/63}{\JournalTitle{\apjs},
  186, 63}

\bibitem[{{Saumon} \& {Marley}(2008)}]{Saumon08}
{Saumon}, D., \& {Marley}, M.~S. 2008,
  \href{http://dx.doi.org/10.1086/592734}{\JournalTitle{\apj}, 689, 1327}

\bibitem[{{Saumon} {et~al.}(2012){Saumon}, {Marley}, {Abel}, {Frommhold}, \&
  {Freedman}}]{Saumon12}
{Saumon}, D., {Marley}, M.~S., {Abel}, M., {Frommhold}, L., \& {Freedman},
  R.~S. 2012,
  \href{http://dx.doi.org/10.1088/0004-637X/750/1/74}{\JournalTitle{\apj}, 750,
  74}

\bibitem[{{Shkolnik} \& {Barman}(2014)}]{Shkolnik14}
{Shkolnik}, E.~L., \& {Barman}, T.~S. 2014,
  \href{http://dx.doi.org/10.1088/0004-6256/148/4/64}{\JournalTitle{\aj}, 148,
  64}

\bibitem[{{Siverd} {et~al.}(2012){Siverd}, {Beatty}, {Pepper}, {Eastman},
  {Collins}, {Bieryla}, {Latham}, {Buchhave}, {Jensen}, {Crepp}, {Street},
  {Stassun}, {Gaudi}, {Berlind}, {Calkins}, {DePoy}, {Esquerdo}, {Fulton},
  {F{\H u}r{\'e}sz}, {Geary}, {Gould}, {Hebb}, {Kielkopf}, {Marshall}, {Pogge},
  {Stanek}, {Stefanik}, {Szentgyorgyi}, {Trueblood}, {Trueblood}, {Stutz}, \&
  {van Saders}}]{Siverd12}
{Siverd}, R.~J., {Beatty}, T.~G., {Pepper}, J., {et~al.} 2012,
  \href{http://dx.doi.org/10.1088/0004-637X/761/2/123}{\JournalTitle{\apj},
  761, 123}

\bibitem[{{Skemer} {et~al.}(2015){Skemer}, {Morley}, {Zimmerman}, {Skrutskie},
  {Leisenring}, {Buenzli}, {Bonnefoy}, {Bailey}, {Hinz}, {Defr{\'e}re},
  {Esposito}, {Apai}, {Biller}, {Brandner}, {Close}, {Crepp}, {De Rosa},
  {Desidera}, {Eisner}, {Fortney}, {Freedman}, {Henning}, {Hofmann},
  {Kopytova}, {Lupu}, {Maire}, {Males}, {Marley}, {Morzinski}, {Oza},
  {Patience}, {Rajan}, {Rieke}, {Schertl}, {Schlieder}, {Stone}, {Su}, {Vaz},
  {Visscher}, {Ward-Duong}, {Weigelt}, \& {Woodward}}]{Skemer15}
{Skemer}, A.~J., {Morley}, C.~V., {Zimmerman}, N.~T., {et~al.} 2015,
  \JournalTitle{ArXiv e-prints},
  \href{http://arxiv.org/abs/1511.09183}{{\sffamily arXiv:1511.09183
  [astro-ph.EP]}}

\bibitem[{{Stassun} {et~al.}(2006){Stassun}, {Mathieu}, \&
  {Valenti}}]{Stassun06}
{Stassun}, K.~G., {Mathieu}, R.~D., \& {Valenti}, J.~A. 2006,
  \href{http://dx.doi.org/10.1038/nature04570}{\JournalTitle{\nat}, 440, 311}

\bibitem[{{Stelzer} {et~al.}(2013){Stelzer}, {Marino}, {Micela},
  {L{\'o}pez-Santiago}, \& {Liefke}}]{Stelzer13}
{Stelzer}, B., {Marino}, A., {Micela}, G., {L{\'o}pez-Santiago}, J., \&
  {Liefke}, C. 2013,
  \href{http://dx.doi.org/10.1093/mnras/stt225}{\JournalTitle{\mnras}, 431,
  2063}

\bibitem[{{Triaud} {et~al.}(2013){Triaud}, {Hebb}, {Anderson}, {Cargile},
  {Collier Cameron}, {Doyle}, {Faedi}, {Gillon}, {Gomez Maqueo Chew},
  {Hellier}, {Jehin}, {Maxted}, {Naef}, {Pepe}, {Pollacco}, {Queloz},
  {S{\'e}gransan}, {Smalley}, {Stassun}, {Udry}, \& {West}}]{Triaud13}
{Triaud}, A.~H.~M.~J., {Hebb}, L., {Anderson}, D.~R., {et~al.} 2013,
  \href{http://dx.doi.org/10.1051/0004-6361/201219643}{\JournalTitle{\aap},
  549, A18}

\bibitem[{{Werner} {et~al.}(2004){Werner}, {Roellig}, {Low}, {Rieke}, {Rieke},
  {Hoffmann}, {Young}, {Houck}, {Brandl}, {Fazio}, {Hora}, {Gehrz}, {Helou},
  {Soifer}, {Stauffer}, {Keene}, {Eisenhardt}, {Gallagher}, {Gautier}, {Irace},
  {Lawrence}, {Simmons}, {Van Cleve}, {Jura}, {Wright}, \&
  {Cruikshank}}]{Werner04}
{Werner}, M.~W., {Roellig}, T.~L., {Low}, F.~J., {et~al.} 2004,
  \href{http://dx.doi.org/10.1086/422992}{\JournalTitle{\apjs}, 154, 1}

\bibitem[{{West} {et~al.}(2008){West}, {Hawley}, {Bochanski}, {Covey}, {Reid},
  {Dhital}, {Hilton}, \& {Masuda}}]{West08}
{West}, A.~A., {Hawley}, S.~L., {Bochanski}, J.~J., {et~al.} 2008,
  \href{http://dx.doi.org/10.1088/0004-6256/135/3/785}{\JournalTitle{\aj}, 135,
  785}

\end{thebibliography}
\end{document}